\documentclass[%
 reprint,
 superscriptaddress,
 amsmath,amssymb,
 aps,
]{revtex4-2}

\usepackage{graphicx}
\usepackage{dcolumn}
\usepackage{bm}
\usepackage{upgreek}
\usepackage{siunitx}
\DeclareSIUnit\molar{\textsc{m}} 
\DeclareSIUnit\Molar{M}        

\def\Supp{Supplemental Material}

\usepackage{kotex}
\usepackage{amsmath}

\begin{document}

\preprint{APS/123-QED}

\title{Density-Dependent Transition in Bacterial Self-Organization Driven by Confinement and Aerotaxis}

\author{Minjun Kim}
\affiliation{Department of Physics, Ulsan National Institute of Science and Technology, Ulsan, Republic of Korea}

\author{Joonwoo Jeong}
\email{jjeong@unist.ac.kr}
\affiliation{Department of Physics, Ulsan National Institute of Science and Technology, Ulsan, Republic of Korea}
\affiliation{UNIST Center for Soft and Living Matter, Ulsan, Korea}

\date{\today}

\begin{abstract}
We experimentally investigate how aerotactic bacteria, confined within a thin liquid film between two solid substrates, respond to a controlled oxygen gradient. 
We find that the total bacterial number density dictates which mechanism dominates the steady-state spatial distribution: wall accumulation or aerotaxis. 
At low densities, despite receiving oxygen only from one substrate, motile bacteria accumulate at both walls, forming a symmetric distribution. 
In contrast, pronounced aerotactic migration toward the oxygen-supplying wall emerges as the density increases. 
Analyzing the temporal evolution of this bacterial distribution reveals that the aerotactic response is driven by a self-generated oxygen gradient induced by collective respiration. 
Our diffusion--advection model of bacteria and oxygen, accounting for aerotactic migration, hydrodynamic attraction to the walls, and respiration, quantitatively reproduces our experimental observations and provides valuable insights into bacterial self-organization within complex environments.

\end{abstract}

\maketitle

\section{Introduction}

Microorganisms often navigate heterogeneous environments where survival depends on their ability to adapt~\cite{Fenchel2002_heterogeneous_world,Spratt2022_phenotypic_variation, Keegstra2022_ecological_roles,Piskovsky2023_motility_resistance,Porter2025GrowthFormColonies}.
Key strategies involve sensing local environments and migrating toward favorable conditions.
For instance, in response to oxygen gradients, aerobic bacteria may perform aerotaxis by biasing their random-walk motility; they can adjust their tumble rate when moving up an attractant gradient~\cite{Fenchel2002_heterogeneous_world,Sokolov2009EnhancedMixing,Ezhilan2012ChaoticDynamics, Douarche2009EcoliOxygen,Bouvard2022AerotacticResponse, Mazzag2003Model,Menolascina2017LogSensing}.
In addition to chemical signals, microorganisms frequently encounter physical boundaries in both natural and engineered settings.
Near these surfaces, a combination of hydrodynamic interactions and persistent self-propulsion leads to significant cell accumulation~\cite{Berke2008Hydrodynamic, Li2009Accumulation, Molaei2014Tumbling, Sartori2018Wall, Junot2022RunToTumble, Li2025algaeaccumu}.
This wall-trapping effect extends residence times at the surface, which in turn increases the probability of surface adhesion and facilitates the critical early stages of bacterial biofilm formation~\cite{Conrad2012NearSurfaceMotility, Fux2005InfectiousBiofilms, Tuson2013BacteriaSurface}.

Bacteria actively reorganize their populations spatiotemporally, responding to and even shaping their complex surroundings. 
These self-organization phenomena have been studied by varying external stimuli~\cite{Jing2020Rheotaxis, Ramamonjy2022NonlinearPhototaxis,Carrere2023MicrophaseSeparation, Prakash2025AlgaeBacteria, Eisenmann2025_puller_instabilities} and confining geometries~\cite{rothschild_non-random_1963, Berke2008Hydrodynamic,vladescu_filling_2014,wioland_confinement_2013,nishiguchi_engineering_2018,xu_self-organization_2019}, or by designing rectifying structures~\cite{Hulme2008_ratchet_sorting,Lambert2010_collective_escape,Denissenko2012_sperm_microchannel,Coppola2021_curved_ratchets,Anand2024_bacterial_rectification}---approaches that have attracted increasing attention for controlling microbial dynamics. 
Microfluidic devices~\cite{Menolascina2017LogSensing,Adler2012AerotaxisMicrofluidic, Oliveira2016TwitchingChemotaxis, Salek2019, Phan2024PKSBreakdown} and capillary tubes~\cite{Adler1966, Zhulin1996_oxygen_taxis, Bouvard2022AerotacticResponse, Detcheverry2025AerotacticBand} have been employed as experimental platforms to study such self-organization, exemplified by traveling bacterial bands, which often emerge from an interplay between the bacterial consumption of limited resources, such as oxygen or nutrients, and the local depletion caused by an insufficient and asymmetric supply. 
In a similar vein, microbial dynamics in nature-mimicking environments featuring strong confinement have garnered significant interest~\cite{Jin2024_MicrobesPorous, Bhattacharjee2021ChemotaxisPorousMedia, Aleklett2018_BuildYourOwnSoil, Porter2025GrowthFormColonies}.

In this work, we introduce a strongly confined one-dimensional system subject to a controlled oxygen supply to investigate the spatiotemporal distribution of aerotactic bacteria. 
Combining systematic experiments that vary the oxygen gradient and bacterial number density with a diffusion--advection model accounting for bacteria-wall interactions and aerotaxis, we reveal an underlying mechanism behind the density-dependent transition in bacterial self-organization: the competition between wall accumulation and a self-generated oxygen gradient. 
These findings demonstrate how bacterial populations actively modify their chemical environment, leading to emergent collective behaviors that depend critically on population density within confined, resource-limited systems.

\section{Results and Discussion}
\begin{figure}[t]
\centering
\includegraphics[width=\columnwidth]{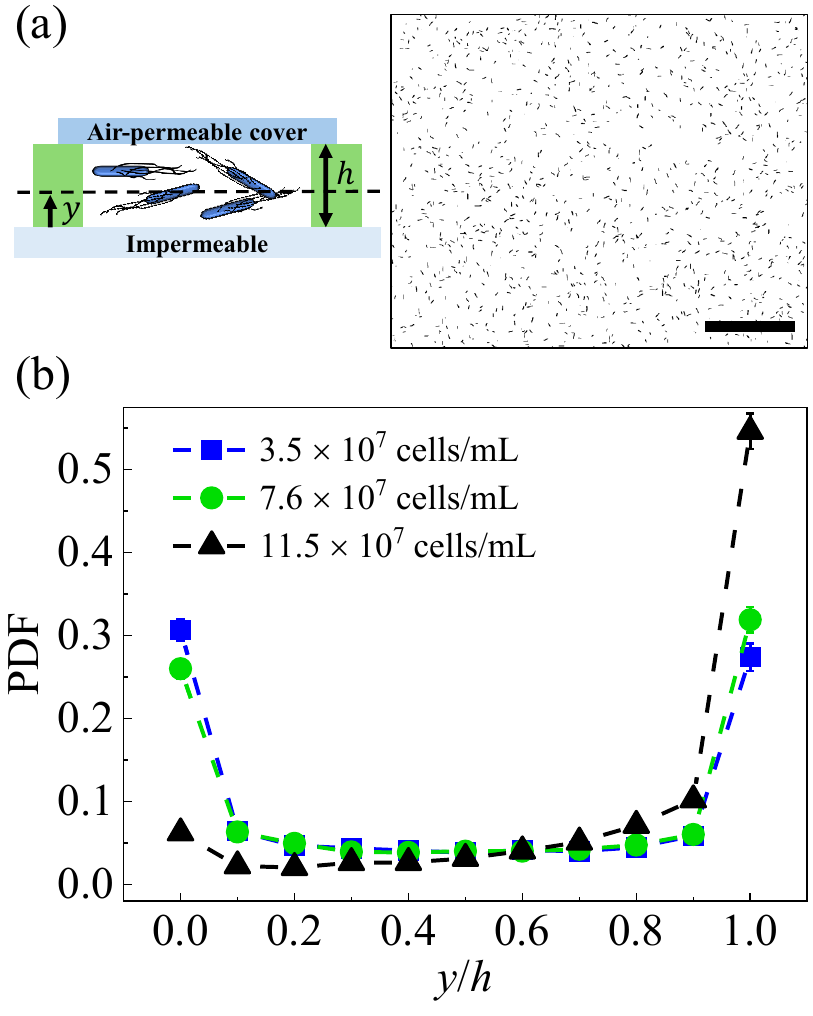}
\caption{
\label{fig:sptial}
Experimental setup and steady-state bacterial distributions for different number densities.
(a) Left: A schematic illustrating the confinement geometry. 
Bacteria are confined in a thin liquid film of height $h$ between a top oxygen-permeable coverslip and a bottom impermeable substrate.
Right: A representative binary image of bacteria at high density near the oxygen-permeable top boundary.
The scale bar is $100$ \si{\um}.
(b) Probability distribution functions (PDFs) of bacterial counts at steady state for three different number densities
The PDFs are plotted against the normalized vertical position, \(y/h\). 
The error bars represent the standard deviation of the temporal fluctuations in the bacterial count at each position, calculated over the recording period.
}
\end{figure}

We investigate the spatial distribution of motile bacteria in a thin liquid film as a function of total bacterial density.
Aerotactic bacteria, \textit{Bacillus subtilis}, are confined in a cylindrical well with an $80–100$ \si{\um} height and $8$ \si {\mm} diameter. 
As shown in Fig.~\ref{fig:sptial}(a), the top boundary is covered with an oxygen-permeable coverslip, while all the other boundaries of the cylindrical well are oxygen-impermeable; the oxygen is supplied exclusively to the top interface.
We divide the film into 11 vertical sections, including the top and bottom boundaries, count the number of swimming bacteria in each section, and measure the one-dimensional distribution function of swimming bacteria; see Materials and Methods for details.

At low number densities such as $3.5\times10^7$ \si{cells/\mL} (Fig.~\ref{fig:sptial}(b), blue squares), we observe a symmetric bacterial distribution along the vertical axis, with population peaks at both top and bottom boundaries: wall accumulation.
The symmetric profile in confinement is primarily attributed to the interaction between motile bacteria and solid boundaries, driven by two key mechanisms: directional persistence and hydrodynamic interactions with boundaries~\cite{Berke2008Hydrodynamic, Li2009Accumulation}.
The former describes the tendency of bacteria to maintain their swimming direction upon encountering an interface, whereas the latter arises from the flow field generated by the rotating flagella modeled as the hydrodynamic dipoles.
Pusher-type swimmers, such as the observed \textit{B. subtilis}, exhibit an effective attraction toward the boundary; consequently, the wall accumulation~\cite{Berke2008Hydrodynamic, Spagnolie2012HydrodynamicsBoundary}.

As the bacterial number density increases, the spatial distributions in Fig.~\ref{fig:sptial}(b) change from symmetric to asymmetric, featuring a pronounced peak near the top oxygen-supplied boundary and a sparse population at the bottom.
Given that oxygen is supplied exclusively from the top interface, this asymmetry is attributed to aerotaxis strong enough to overcome the wall accumulation effect.
A series of control experiments, including one with a gas-controlled chamber (Sup. Fig. S1 and S2), confirms that aerotaxis is the primary driver of the observed asymmetric organization.
Based on these findings, the total bacterial number density $B_{\mathrm{tot}}$ is the key parameter that determines the balance between wall accumulation and aerotaxis in this confined, resource-limited environment.
Notably, a supplementary experiment with induced spatial inhomogeneity (Sup. Fig. S3) demonstrates that the steady-state distribution is set by the total population and global oxygen balance rather than local density variations.

\begin{figure*}
\centering
\includegraphics[width=\textwidth]{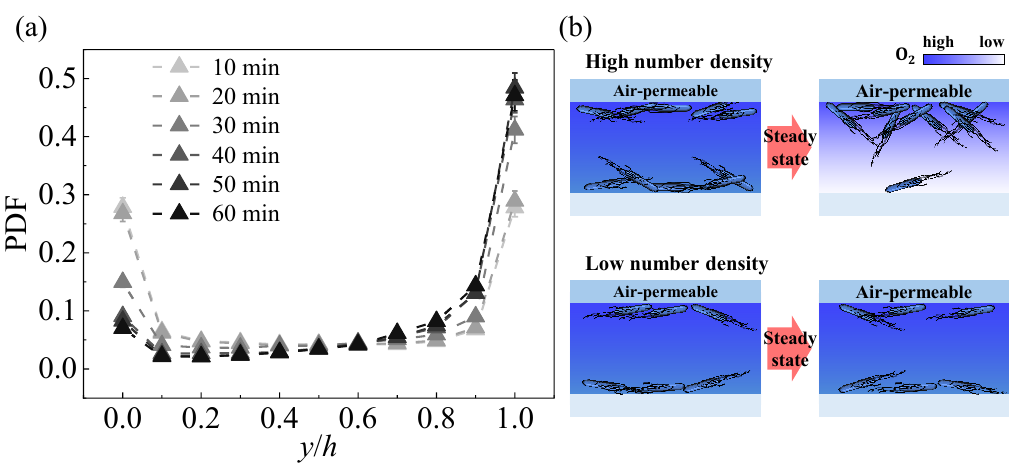}
\caption{
\label{fig:temporal}
Temporal evolution of the bacterial distribution at high density and a schematic of the system dynamics.
(a) PDFs of bacterial counts for a high-density suspension ($11.7\times10^7$ \si{cells/\mL}) measured at different time points, from 10 min (light gray triangles) to 60 min (black triangles).
The error bars represent the standard deviation of temporal fluctuations in the bacterial count at each position, calculated over the recording period.
(b) Schematic diagrams contrasting the system's evolution at high and low number densities. 
The top row depicts the high-density case, showing a transition from an initial symmetric distribution to an asymmetric steady state.
The bottom row depicts the low-density case, showing the persistence of a symmetric distribution from the initial to the steady state.
The color fill in the diagrams represents the oxygen concentration, which is defined by the legend in the top-right corner (labeled $\mathrm{O}_2$); blue and white indicate high and low concentrations, respectively.
}
\end{figure*}

Temporal evolution measurements in high-density suspensions ($11.7 \times 10^7$ \si{cells/\mL}) suggest a self-generated oxygen gradient as the mechanism underlying the total density-dependent spatial distribution.
As shown in Fig.~\ref{fig:temporal}(a), an initially symmetric, wall-accumulation-dominated distribution transitions to an asymmetric steady state governed by aerotactic migration as time goes by.
In contrast, low-density suspensions maintain their symmetric distribution throughout the observation period (Sup. Fig. S4), indicating that significant collective respiration is required to generate oxygen depletion and vertical gradients, thus breaking the spatial symmetry.
This observation is consistent with previous studies showing that motile bacteria in confinement create self-organized patterns by consuming local resources~\cite{Bouvard2022AerotacticResponse, Phan2024PKSBreakdown, Hokmabad2025SelfOrganization}.

Synthesizing these findings, we propose a conceptual model to explain the observation (Fig.~\ref{fig:temporal}(b)).
The symmetric, wall-dominated distribution develops when the oxygen concentration is uniform. 
As respiration proceeds, oxygen is preferentially depleted near the oxygen-impermeable interface, establishing an oxygen-concentration gradient.
In high-density suspensions, where the collective consumption rate exceeds the oxygen-supply flux from the oxygen-permeable interface, this self-generated gradient drives a migration toward the oxygen-rich region via aerotaxis, resulting in an asymmetric steady-state profile. 
At low density, however, the respiration rate remains below the replenishment rate, and the bacterial population does not experience a strong enough oxygen gradient, so it maintains a symmetric distribution.
Ultimately, the total bacterial number density determines whether the system is governed by wall accumulation or by emergent, self-generated oxygen gradients.

To rationalize our experimental results, we introduce a continuum model based on the Keller--Segel framework, which describes the coupled dynamics of motile bacteria and a chemo-attractant (oxygen)~\cite{Keller1971ModelChemotaxis, Keller1971TravelingBands}.
The following diffusion--advection equation describes the spatiotemporal evolution of the one-dimensional bacterial density, \(B(y,t)\) along the spatial axis ($y$):
\begin{equation}
  \frac{\partial B(y,t)}{\partial t}=\frac{\partial}{\partial y}[D_{\mathrm{bac}}\frac{\partial B(y,t)}{\partial y}-u(y,t)B(y,t)],
  \label{eq:bac_total}
\end{equation}
where \(D_{\mathrm{bac}}\) is the bacterial diffusion coefficient representing random motility of bacteria, and \(u(y,t)\) is the directional migration velocity.
Our justification for applying the Keller--Segel framework to a confined system, where the film thickness ($h\approx\SI{100}{\um}$) is not significantly larger than the bacterial run length of approximately \SI{20}{\um}, is detailed in the \Supp.

The migration velocity $u(y,t)$ to account for wall accumulation and aerotaxis in our model comprises hydrodynamic velocity (\(u_{\mathrm{hydro}}\)) and aerotactic velocity (\(u_{\mathrm{taxis}}\)):
\begin{equation}
    u(y,t)=u_{\mathrm{hydro}}+u_{\mathrm{taxis}}.
    \label{eq:bac_adv}
\end{equation}
The hydrodynamic term describes the attraction of pusher-type swimmers to nearby surfaces, resulting in the wall accumulation, and the aerotactic term represents the biased migration up the oxygen concentration gradient:
\begin{align}
  u_{\mathrm{hydro}}=-\frac{3p}{64\pi\eta}(\frac{1}{y^2}-\frac{1}{(h-y)^2})
  \label{eq:bac_hydro}
  \\
  u_{\mathrm{taxis}}=\chi\frac{\partial c(y,t)/\partial y}{(c(y,t)+K_{\mathrm{d1}})(c(y,t)+K_{\mathrm{d2}})},
  \label{eq:bac_taxis}
\end{align}
where \(h\) is the sample height, \(p\) is the dipole strength of bacteria, \(\eta\) is viscosity, \(\chi\) is aerotactic sensitivity, \(K_{\mathrm{d1}}\), \(K_{\mathrm{d2}}\) are oxygen dissociation constants of \textit{B. subtilis}, and \(c(y,t)\) is local oxygen concentration~\cite{Berke2008Hydrodynamic, Menolascina2017LogSensing, Zhang2005HemAT, Tu2008Chemotaxismodel}. 
To avoid the divergence of \(u_{\mathrm{hydro}}\) at the boundaries ($y=0$ and $y=h$), we adopt a regularized Stokeslet formulation, replacing \(\frac{1}{y^2}\) and \(\frac{1}{(h-y)^2}\) with \(\frac{1}{y^2+\epsilon^2}\) and \(\frac{1}{(h-y)^2+\epsilon^2}\), where $\epsilon$ is the blob size~\cite{Ainley2008RegularizedImages, Cisneros2008BipolarFlows, Zhao2019RegularizedStokeslets, Prabhune2024TugOfOars} (details in \Supp).
Note that this model with $u_{\mathrm{hydro}}$ considers the wall accumulation only by the hydrodynamic interaction, not by the directional persistence of bacteria. 
\Supp~discusses this limitation further and future works.

The oxygen concentration \(c(y,t)\), which is governed by the oxygen diffusion and its consumption by bacteria, is coupled to the bacterial density \(B(y,t)\) via \(u_{\mathrm{taxis}}\). \(c(y,t)\) follows
\begin{equation}
  \frac{\partial c(y,t)}{\partial t}=D_{\mathrm{oxy}}\frac{\partial^2 c(y,t)}{\partial y^2}-R(y,t)B(y,t)
  \label{eq:oxy_total}
\end{equation}
where \(D_{\mathrm{oxy}}\) is the diffusion coefficient of oxygen dissolved in water and \(R(y,t)=\frac{R_0}{2}(\frac{c(y,t)}{c(y,t)+K_{\mathrm{d1}}}+\frac{c(y,t)}{c(y,t)+K_{\mathrm{d2}}})\) is the oxygen respiration rate per bacterium, which depends on the local oxygen concentration~\cite{Zhang2005HemAT}.


\begin{figure}[t]
\centering
\includegraphics{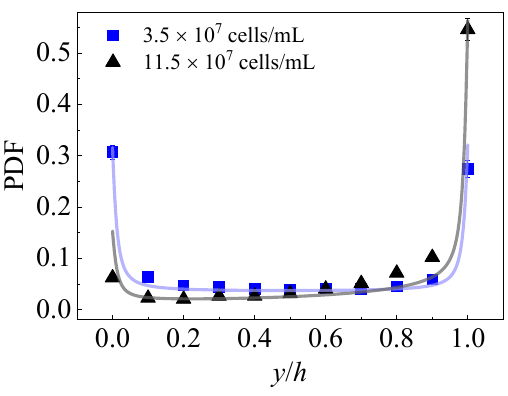}
\caption{
\label{fig:model fit}
Comparison between experimental bacterial distributions (symbols) and model predictions (solid lines) for low and high number densities.
Two representative bacterial number densities in Fig.~\ref{fig:sptial} are chosen for numerical fitting.
}
\end{figure}


We numerically solve two coupled, nondimensionalized partial differential equations for the steady state subject to appropriate boundary conditions and, as shown in Fig.~\ref{fig:model fit}, reproduce the density-dependent spatial distributions observed in the experiments.
See Materials and Methods for the calculation details.
We fit the experimental data numerically by minimizing the mean squared error (MSE) across the three densities and find all fitted parameters are within the range of previously reported values~\cite{Zhang2005HemAT, Menolascina2017LogSensing, Tuval2005ContactLines,Cheon2024MotileInterface}: $\chi = 2.0\times10^{3}$ \si{\Molar~\um^{2}/\s}, $K_{\mathrm{d1}}$ = \SI{1.0}{\micro\Molar}, $K_{\mathrm{d2}}$ = \SI{100}{\micro\Molar}, $k=\frac{3p}{64\pi\eta} = 6.0\times10^{2}$ \si{\um^{3}/\s}, $R_{0} = 2.0\times10^{6}$ \si{molecules/\s} per bacterium, and $\epsilon$ = \SI{2}{\micro m}.
Here, the blob size includes the radius of the cell body and the flagella bundle.
Fluorescence images of the flagellar bundle show a maximum width of $\sim $ \SI{7}{\micro m}, thus $\epsilon$ = \SI{2}{\micro m} lies within this range~\cite{Najafi2019MultipleBundles}.
Note also that we adopt the diffusion coefficients for bacteria and oxygen from reference~\cite{Menolascina2017LogSensing}, $D_{\mathrm{bac}} = 2.1\times10^{2}$ \si{\um^{2}/\s}, $D_{\mathrm{oxy}} = 2\times10^{3}$ \si{\um^{2}/\s}.


\begin{figure}
\centering
\includegraphics[width=0.5\textwidth]{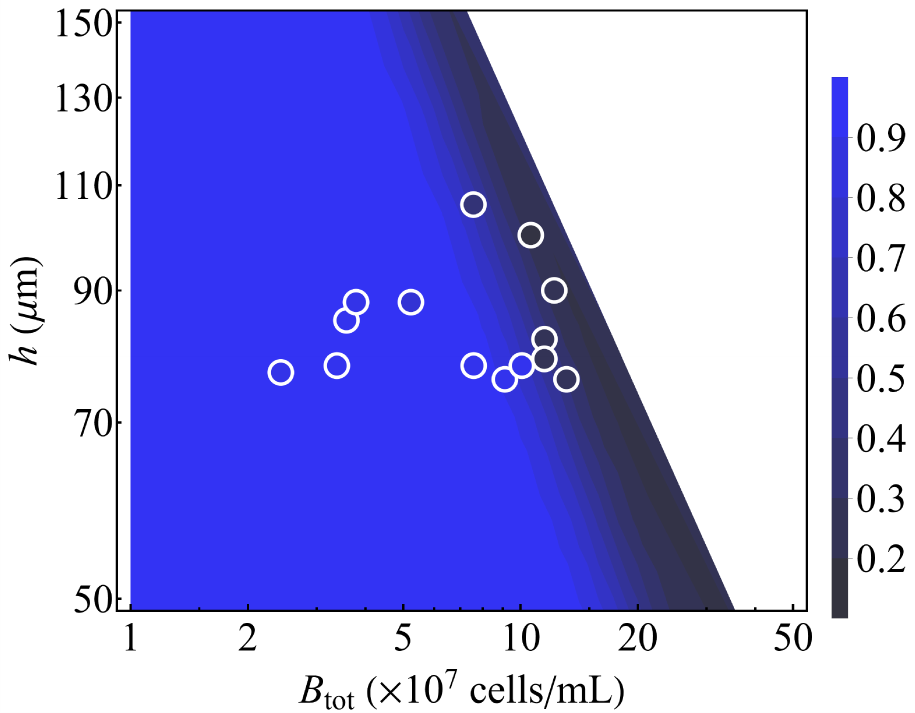}
\caption{
\label{fig:state diagram}
The state diagram of bacterial distribution symmetry as a function of total bacterial number density (\(B_{\mathrm{tot}}\)) and film height ($h$).
The color scale shows the symmetry parameter ($sym$); darker colors indicate more asymmetric (aerotaxis-dominated) profiles.
Regions of $sym<0.17$ are omitted because the numerical solutions fail to satisfy the boundary conditions in this highly asymmetric regime.
Experimental data are overlaid as circles which are color-coded with the experimentally estimated $sym$.
}
\end{figure}


Beyond fitting the experimental data, we use our model to explore the parameter space defined by two primary control parameters, the total bacterial number density ($B_\mathrm{tot}$) and film height ($h$).
First, to quantify how symmetric the bacterial distribution is, we define a symmetry parameter \[sym=\frac{\mathrm{min}(L_\mathrm{S},U_\mathrm{S})}{\mathrm{max}(L_\mathrm{S},U_\mathrm{S})}\] where \(L_\mathrm{S}=\int_0^{h/2} B(y)\,\mathrm{d}y\) and \(U_\mathrm{S}=\int_{h/2}^h B(y)\,\mathrm{d}y\).
Thus, $sym=1$ indicates perfect symmetry, whereas smaller values indicate stronger asymmetry.
As shown in Fig.~\ref{fig:state diagram}, we then calculate this symmetry parameter over a range of $B_\mathrm{tot}$ and $h$ to generate a state diagram overlaid with experimental data.
As explained, as the cell density $B_\mathrm{tot}$ increases, $sym$ decreases because of the self-generated oxygen gradient.
The dependence on film height ($h$) can be understood through the oxygen diffusion time, \(\tau_{\mathrm{oxy}}=h^2/D_{\mathrm{oxy}}\).
As the film gets thinner, the diffusion time gets shorter, meaning oxygen supplied from the top boundary can diffuse faster throughout the entire film.
This rapid replenishment prevents the formation of strong depletion, thus weakening the aerotactic response and favoring a symmetric distribution.
Our model calculation can be extended to a broader range of \(B_{\mathrm{tot}}\) and \(h\), including the bioconvection regime~\cite{Yanaoka2022Bioconvection3D, GallardoNavarro2025Segregation, Shoup2023Bioconvection} and extreme confinement effects~\cite{Wei2025Confinement}, as detailed in the \Supp.

\section{Conclusion}
In conclusion, we have investigated bacterial self-organization under confinement subject to an inhomogeneous oxygen supply. 
The total bacterial number density dictates the dominant mechanism governing their spatial distribution: wall accumulation prevails at low densities, whereas a self-generated oxygen gradient, driven by collective bacterial respiration, induces aerotactic migration at high densities, leading to an asymmetric distribution. 
Our continuum model, combining the Keller–-Segel framework with the hydrodynamic attraction of bacteria to surfaces, quantitatively reproduces this density-dependent transition across a broad range of bacterial densities and film heights.

Our work establishes a framework for understanding how bacterial populations respond to complex environments where chemical stimuli and physical confinement coexist. 
Given that such conditions --- where bacteria simultaneously experience spatial restriction and resource limitation --- are prevalent in their natural habitats, ranging from soil matrices to mucus layers, a predictive model of their combined effects is crucial for advancing our understanding of bacterial organization. 
Specifically, our findings offer insights into how confined bacteria not only respond to but also collectively shape their chemical environment, with potential implications for biofilm formation and survival strategies under inhospitable conditions. 
Furthermore, the density-dependent transition we observe may reflect a more general principle applicable across diverse spatiotemporal scales, such as animal populations competing for limited resources in spatially restricted habitats.

\section{Materials and Methods}

\subsection{Bacterial cell culture}
A single colony of \textit{Bacillus subtilis} strain ATCC 6051 (wild type) was initially inoculated onto Lysogeny Broth (LB) agar and incubated overnight at \SI{37}{\degreeCelsius}. 
The resulting culture was then transferred to Terrific Broth (TB) and grown overnight in a shaking incubator at \SI{37}{\degreeCelsius} and 250 RPM.
This diluted culture ---  with fresh TB to an optical density at 600 \si{\nm} (OD600) of approximately 0.03 --- was then incubated under the same conditions (\SI{37}{\degreeCelsius}, 250 RPM) until it reached the mid-exponential phase.
To maximize cell yield, the culture was transferred to \SI{15}{\ml} tubes and centrifuged at 4000 RPM for \SI{15}{\min}, then resuspended in a final volume of \SI{2}{\ml} TB. 
To predominantly single, motile bacteria with minimal aggregated bacteria, the \SI{1}{\ml} solution was centrifuged at 6000 RPM for \SI{1}{\min}, and \SI{800}{\ul} of the supernatant was carefully collected.
Finally, this bacterial suspension was washed and resuspended in M9 minimal medium to reduce cell division during the observation.

\subsection{Sample preparation}
To construct the observation chamber, we sandwiched two substrates using double-sided aluminum tape (No. 791, Teraoka, Japan) as a spacer, which had an \SI{8}{\mm} diameter made by a biopsy punch (Miltex, Ted Pella, U.S.A).
We confirmed that the spacer and a glass substrate functioned as an oxygen barrier, while the other substrate, the coverslip (Cat. no. 10813, Ibidi, Germany), is oxygen-permeable.
The drop-and-cover assembly with approximately \SI{8.0}{\ul} of the bacterial suspension resulted in a confined liquid film with an approximate height of 80-\SI{100}{\um}.
Alternatively, in the number fluctuation experiment, the chamber was filled with bacterial suspension via capillary flow, i.e., instead of the drop-and-cover method, and sealed with epoxy (Loctite, Henkel, Germany)~\cite{Bodhankar2022phenol, Henkel2025epoxysds}.
To generate a radial oxygen supply, we also used a PDMS-based sheet (SecureSeal, Grace Bio-Labs, U.S.A.) of \SI{120}{\um} thickness as an oxygen-permeable spacer.
Experiments controlling the external oxygen concentration were conducted in a live-cell imaging chamber equipped with a gas-mixing chamber (Live Cell Instrument, Republic of Korea), which provided the chamber with a mixed gas of ambient air and nitrogen at a controlled ratio.

\subsection{Optical microscopy and measurement}
We observed bacteria using an inverted microscope (IX73, Olympus) equipped with 20X and 40X objective lenses and a CCD camera (INFINITY5, Teledyne Lumenera, Canada) in both bright-field and phase-contrast modes.
To determine the bacterial distribution along the sample height, we counted bacteria at 11 distinct, equally spaced positions.
This imaging range extended from the bottom impermeable boundary to the oxygen-permeable top cover of the sample. 
At each of these 11 positions, we recorded a 7-to-15-second-long movie at 30-50 frames per second.

To segment focused bacteria at each focal plane, we adopted Statistical Regional Merging (SRM)~\cite{Nock2004SRM} and applied an appropriate threshold to the SRM-processed data.
Note that we counted only motile bacteria, to account for the wall accumulation effect, excluding sedimented cells of which displacements over 1.8 sec were less than one body length ($\sim10$ \si{\um}).
We found that over 85\% of the total population were motile.

\subsection{Model calculation}
We performed numerical calculations for our theoretical model using Mathematica 14.3 (Wolfram). 
The steady-state distributions of bacteria and oxygen were obtained by solving Eq.~\ref{eq:bac_total} and \ref{eq:oxy_total} satisfying \(\int_0^h B(y)\,\mathrm{d}y=B_{\mathrm{tot}}\).
The equations were nondimensionalized as shown in the \Supp~and solved numerically with the following boundary conditions.
For bacteria, no-flux boundary conditions are imposed at both surfaces (\(y=0,h\)).
For oxygen, the bottom surface (\(y=0\)) is also a no-flux boundary, while the oxygen can be supplied from the top surface (\(y=h\)).
Avoiding numerical instability, we implement a Dirichlet boundary condition at the top, \(c(y=h)=c_h\), and determine the value of $c_h$ self-consistently.
Namely, at the steady-state with $c_h$, the total oxygen flux entering through the top boundary, $J_{\mathrm{oxy}}(c_h)$, equals the total consumption rate by all bacteria in the film, $R_\mathrm{tot}(c_h)$.
Specifically, at the steady state, the oxygen flux was given by \(J_{\mathrm{oxy}}=-D_{\mathrm{cover}}\frac{c_{\mathrm{air}}-c_{h}}{h_{\mathrm{cover}}}\), where $D_{\mathrm{cover}}$ was the oxygen diffusion coefficient in the oxygen-permeable coverslip, $c_{\mathrm{air}}$ is the oxygen concentration in ambient air, and $h_{\mathrm{cover}}$ is the thickness of the coverslip. 
We adopted \(D_{\mathrm{cover}} = 3.19\times10^{-1}\) \si{\um^{2}/\s} and \(h_{\mathrm{cover}} = 180 \) \si{\um} and searched $c_h$ that satisfied the steady-state condition with the range of \(5\times10^{-5} c_{0}\) and \(c_{0}\), where $c_0$ denotes the saturated oxygen concentration in water.
When the computed value exceeded $c_0$, we set \(c_h=c_{0}\).
We found that all \(c_h\) values in our calculations were above the motility threshold, at which cells lose their motility at \(c_h<5\times10^{-5} c_{0}\)~\cite{Hokmabad2025SelfOrganization}.
Lastly, we do not show extremely asymmetric solutions ($sym<0.17$) in Fig.~\ref{fig:state diagram}, as they often fail to satisfy the boundary conditions numerically.


%

\end{document}


\preprint{APS/123-QED}

\title{Supplemental Materials for ``Density-Dependent Transition in Bacterial Self-Organization Driven by Confinement and Aerotaxis''}

\author{Minjun Kim}
\affiliation{Department of Physics, Ulsan National Institute of Science and Technology, Ulsan, Republic of Korea}

\author{Joonwoo Jeong}
\email{jjeong@unist.ac.kr}
\affiliation{Department of Physics, Ulsan National Institute of Science and Technology, Ulsan, Republic of Korea}
\affiliation{UNIST Center for Soft and Living Matter, Ulsan, Korea}

\date{\today}

\maketitle


\section{Control experiments verifying aerotactic migrations}

\begin{figure*}
\centering
\includegraphics[width=\textwidth]{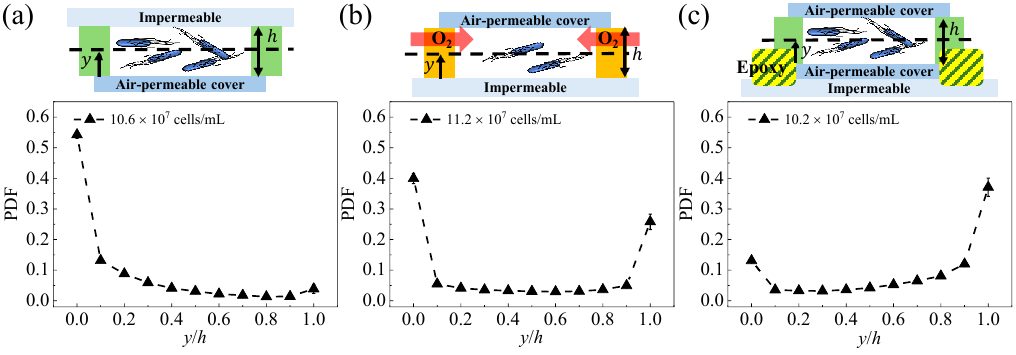}
\caption{
\label{fig:control}
Control experiments to test the aerotaxis hypothesis. The top row shows the schematics, and the bottom row shows the steady-state bacterial distribution as a function of normalized vertical position.
(a) Flipped film. (b) Oxygen-permeable spacer. (c) Surface affinity check.
The x-axis represents the normalized vertical position ($y/h$), and error bars indicate temporal fluctuation.
}
\end{figure*}

\begin{figure*}
\centering
\includegraphics[width=0.49\columnwidth]{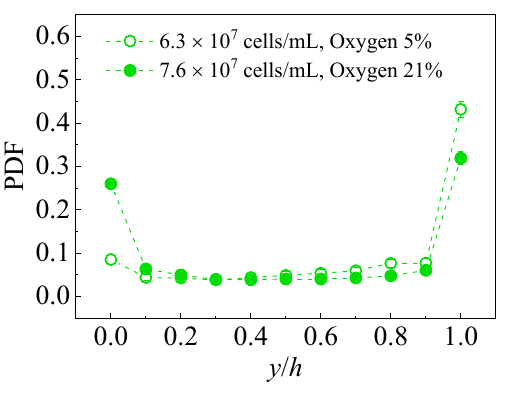}
\caption{
\label{fig:chamber}
Bacterial distributions measured in an oxygen concentration-controlled gas chamber.
The experiment compares a hypoxia condition (5\% O$_2$, $6.3 \times 10^7$ cells/mL; open circles) with an ambient-air condition (21\% O$_2$, $7.6 \times 10^7$ cells/mL; filled circles). 
The x-axis represents the normalized vertical position ($y/h$), and error bars indicate temporal fluctuation, as in Fig.~\ref{fig:control}.
}
\end{figure*}

To confirm that the aerotaxis drives the asymmetric distribution at high number densities, we conduct the following control experiments (Fig.~\ref{fig:control}).
First, as shown in Fig.~\ref{fig:control}(a), we flip the entire liquid film with respect to the gravitational direction and find that the bacterial accumulation shifts toward the new oxygen-supplying boundary at $y/h=0$.
This observation assures that the oxygen-supplying boundary matters in the asymmetric distribution, disputing any possible role of gravity. 
Second, Fig.~\ref{fig:control}(b) shows that the oxygen-permeable spacer eliminates the asymmetric profile even under high $B_{\mathrm{tot}}$, supporting our hypothesis of self-generated oxygen gradient under limited oxygen supply.
Third, as shown in Fig.~\ref{fig:control}(c), we demonstrate that oxygen-permeable cover with no access to oxygen does not attract bacteria, excluding a surface affinity scenario that bacteria like the surface of the oxygen-permeable covers, not the oxygen supplied from them.
The degassed, oxygen-permeable cover was stored in a nitrogen chamber overnight prior to the film assembly.
As shown in the schematic, we block its exposure to air by gluing a glass substrate to the cover with epoxy and observe an asymmetric bacterial distribution.

Lastly, we conduct gas-chamber experiments, controlling the oxygen concentration in the gas chamber (Fig.~\ref{fig:chamber}).
When a sample with the intermediate number density ($B_{\mathrm{tot}}=6.3 \times 10^7$ cells/\si{\mL}) is exposed to an external oxygen concentration of 5\%, an asymmetric distribution is observed.
Note that, as shown in Fig.~\ref{fig:chamber}, the sample with a slightly higher bacterial density ($B_{\mathrm{tot}}=7.6 \times 10^7$ cells/\si{\mL}) exhibits a symmetric distribution in the ambient condition (21\% oxygen).
These observations nail down that the migration is oxygen-driven.

\FloatBarrier

\section{Bacterial distribution is dictated by global density rather than local inhomogeneities}

\begin{figure*}
\centering
\includegraphics[width=\textwidth]{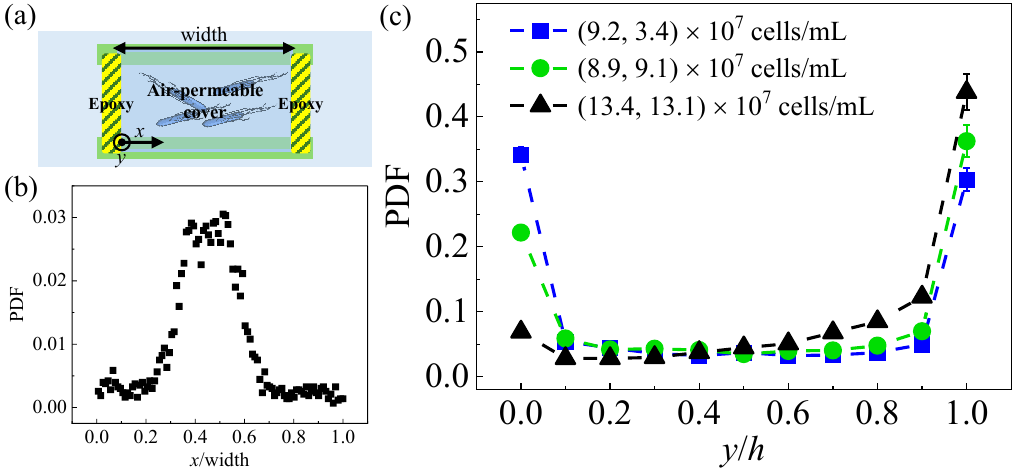}
\caption{
\label{fig:heterogeneity}
Set up for the spatial heterogeneity experiment and resulting bacterial distributions at different local number densities.
(a) Schematic of the experimental setup. 
The bacterial suspension is injected into the narrow gap by capillary flow.
The entrance and exit were subsequently sealed with epoxy to minimize water evaporation and induce negative chemotaxis, i.e., bacteria repel from the epoxy.
The position from the entrance is denoted by $x$, and the width is the distance between the two epoxy boundaries.
(b) Spatial inhomogeneity in the low cell density suspension.
The bacterial population is measured along the horizontal ($x$) direction, and the PDF is plotted as a function of the normalized position ($x/\mathrm{width}$).
(c) Steady-state distributions of bacteria measured at the central region for different total densities: blue squares (low), green circles (intermediate), and black triangles (high).
The legend values denote $(B_{\mathrm{local}}, B_{\mathrm{tot}})$ representing the local number density at the center and the total density for each suspension, respectively.
}
\end{figure*}

We identify the total number density ($B_{\mathrm{tot}}$) as the key parameter determining the dominant mechanism, i.e., wall accumulation versus aerotaxis, in our confined liquid film system. 
However, local density fluctuations can potentially alter the local oxygen concentration, leading to spatially varying distributions across the film. 
In fact, such density heterogeneity has been reported in free-standing liquid films thicker than 200 \si{\um}~\cite{Sokolov2009EnhancedMixing, Ezhilan2012ChaoticDynamics}. 
To investigate this, we intentionally induce spatial inhomogeneity within a confined film by sealing the confined bacterial suspension using epoxy (Fig.~\ref{fig:heterogeneity}(a)), instead of the double-sided aluminum tape used in our typical experiments. 
For three samples with different total bacterial densities, we measure the bacterial distributions along the vertical direction $y$ at the film center along $x$.
Phenol, present in the epoxy hardener, acts as a chemo-repellent at millimolar concentration~\cite{Bodhankar2022phenol, Henkel2025epoxysds}.
Because we find that the uncured epoxy acts as a chemo-repellent, as shown in Fig.~\ref{fig:heterogeneity}(b), the bacterial density has a peak at the center in low $B_{\mathrm{tot}}$ suspensions.
Steric interactions at higher densities broaden the bacterial accumulation at the center, weakening the spatial heterogeneity.
Notably, in this case, the \textit{local} number density at the center exceeds the local and global densities of the intermediate $B_{\mathrm{tot}}$ case.
However, as shown in Fig.~\ref{fig:heterogeneity}(c), the distribution turns out to be symmetric.
This implies that the symmetry of the bacterial density profile depends strongly on $B_{\mathrm{tot}}$ rather than $B_{\mathrm{local}}$, suggesting that global bacterial respiration and oxygen concentration play crucial roles.

\FloatBarrier

\section{Temporal evolution of a bacterial distribution at a low bacterial density}

\begin{figure*}
\centering
\includegraphics[width=0.45\textwidth]{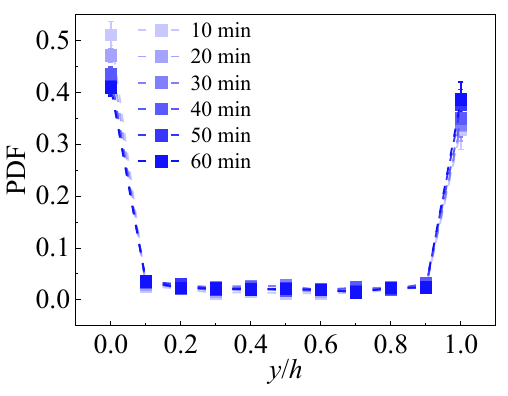}
\caption{
\label{fig:temporal low density}
Temporal evolution of the bacterial distribution at a low density (\(3.8\times10^7\) cells/mL).
}
\end{figure*}

We measure the temporal evolution of the bacterial distribution with a low number density (\(3.8\times10^7\) cells/mL) at 10-min intervals, as shown in Fig.~\ref{fig:temporal low density}.
The distribution remains symmetric and stable throughout the observed period, unlike the high-density case shown in Fig.~2(a) of the main text.

\FloatBarrier

\section{Model justification for a confined system}

First, we address the applicability of our continuum model to the thin-film geometry, specifically where the film height $h$ is comparable to the bacterial run length ($l_{\mathrm{run}}\sim$ \SI{20}{\um}). 
The Keller--Segel framework provides a macroscopic description of bacterial transport under chemical gradients and is formally derived from the Fokker--Planck equation governing underlying run-and-tumble dynamics.
A fundamental assumption for this coarse-graining is that bacteria experience only a marginal fractional change in attractant concentration during a single run~\cite{Mazzag2003Model,Detcheverry2025AerotacticBand}. 
For instance, the Supplemental Material of Menolascina et al.~\cite{Menolascina2017LogSensing} describe this ``shallow gradient'' criterion as follows:
    \begin{equation}
    \frac{v}{c_{\mathrm{adapt}}}\left|\frac{(K_{\mathrm{d1}}-K_{\mathrm{d2}})\partial c(y,t)/\partial y}{(K_{\mathrm{d1}}+c(y,t))(K_{\mathrm{d2}}+c(y,t))}\right| \leq \frac{l_{\mathrm{run}}}{h},
    \label{eq:shallowgrad}
    \end{equation}
where $c(y,t)$ is the attractant concentration field. 
Most parameters are from our experiments and numerical fitting, from $h \approx \SI{100}{\um}$ and oxygen dissociation constants $K_{\text{d1}}$ and $K_{\text{d2}}$ to the swimming velocity $v\sim$ \SI{15}{\um/\s} and the run length $l_{\mathrm{run}}\sim$ \SI{20}{\um}. 
For the adaptation rate $c_{\mathrm{adapt}}$ of the bacteria to chemical stimuli, we adopt $c_{\mathrm{adapt}} \approx 1.4 \times 10^{-2}$~\si{s^{-1}} from established \textit{E. coli} models~\cite{Tu2008Chemotaxismodel}. 
We confirm that our confined system satisfies Eq.~\ref{eq:shallowgrad} within the regime defined by this adaptation rate, although a precise measurement for $c_{\mathrm{adapt}}$ of \textit{B. subtilis} remains beyond the scope of the present work.
In fact, the same report~\cite{Menolascina2017LogSensing} successfully described their experimental data having various oxygen gradients with the Keller-Segel framework, and our calculated oxygen gradient --- approximately $0.0002~\%~\mathrm{\mu m^{-1}}$ when $100~\%$ denotes the saturation concentration at room temperature --- falls within the investigated gradient range.
Although the Keller-Segel model seems to be an effective, minimal model to rationalize our steady-state observation of aerotactic bacteria distribution, we should note a caveat that steep gradients near concentrated bacterial bands can occasionally challenge the exact application of this model~\cite{Phan2024PKSBreakdown, Detcheverry2025AerotacticBand}.

Second, we discuss the incorporation of the wall accumulation into the Keller--Segel framework. 
We employ hydrodynamic interactions as the primary mechanism for wall accumulation, a choice motivated by its seamless integration into our diffusion--advection framework.
However, it is important to acknowledge an alternative origin of wall accumulation: the persistent self-propulsion of bacteria toward solid surfaces~\cite{Li2009Accumulation, Sartori2018Wall}.
Near a boundary, wall-parallel reorientation and suppressed tumbling reduce the escape probability, thereby enhancing accumulation at the boundaries~\cite{Li2009Accumulation, Molaei2014Tumbling}.
Since our model is formulated within a one-dimensional framework, rotational diffusion does not appear explicitly in our equations.
Nevertheless, its effective contribution could be incorporated by introducing a position-dependent translational diffusion coefficient $D_{\mathrm{bac}}(y)$ that decreases near the boundaries to capture the effects of reduced tumbling and wall-parallel alignment.
Such an extension, which would unify hydrodynamic and self-propulsion effects within a single framework, remains for future work.

\section{Application of regularized Stokeslet in our model}

To avoid divergence at the top and bottom boundaries of our model, we employ the regularized Stokeslet for $u_{\mathrm{hydro}}=-k(\frac{1}{y^2}-\frac{1}{(h-y)^2})$~\cite{Berke2008Hydrodynamic}.
Starting from the expression of a normal Stokeslet for a dipole, the flow velocity becomes
\begin{equation}
    u^{\mathrm{dipole}}(\bm{r})=\frac{1}{4\pi\eta}(\frac{\bm{q}}{r^3}-\frac{3(\bm{q}\cdot\bm{r})\bm{r}}{r^5}),
    \label{eq:dipole}
\end{equation}
where $\bm{q}$ is the dipole moment~\cite{Ainley2008RegularizedImages}. 
With the general blob function $\phi_{\epsilon}(r)=\frac{15\epsilon^4}{8\pi(r^2+\epsilon^2)^{7/2}}$~\cite{Ainley2008RegularizedImages, Zhao2019RegularizedStokeslets}, the regularized Stokeslet becomes
\begin{equation}
    u_{\epsilon}^{\mathrm{dipole}}(\bm{r})=\frac{1}{4\pi\eta}(\frac{\bm{q}}{(r^2+\epsilon^2)^{3/2}}-\frac{3(\bm{q}\cdot\bm{r})\bm{r}}{(r^2+\epsilon^2)^{5/2}})-\frac{3\epsilon^2\bm{q}}{(r^2+\epsilon^2)^{5/2}} 
    \approx \frac{1}{4\pi\eta}(\frac{\bm{q}}{(r^2+\epsilon^2)^{3/2}}-\frac{3(\bm{q}\cdot\bm{r})\bm{r}}{(r^2+\epsilon^2)^{5/2}}).
    \label{eq:dipole_reg_deriv}
\end{equation}
Comparing Eq.~\ref{eq:dipole} with \ref{eq:dipole_reg_deriv}, we find that $r$ is replaced to be $\sqrt{r^2+\epsilon^2}$, and modify the hydrodynamic advection consequently:
\begin{equation}
    u_{\mathrm{hydro}}=-k(\frac{1}{y^2+\epsilon^2}-\frac{1}{(h-y)^2+\epsilon^2}).
    \label{eq:dipole_reg}
\end{equation}

\section{Nondimensionalization of model equations}

Here we nondimensionalize Eq. 1 and 5 of the main text.
First, we define nondimensionalized variables: $\eta=y/h$ and $\tau=t/\tau_0$, where $h$ is the confinement height and $\tau_0=h^2/D_{\mathrm{oxy}}$ as the characteristic diffusion time of oxygen, respectively. 
Then, $\theta(\eta,\tau)=c(y,t)/c_0$ and $\xi(\eta,\tau)=B(y,t)/B_{\mathrm{tot}}$ represent the nondimensionalized oxygen concentration and bacterial density.
$c_0$ is the saturated oxygen concentration in water at room temperature, and $B_{\mathrm{tot}}$ is the total bacteria number density.
Equation 5 becomes
\begin{equation}
  \frac{\partial \theta(\eta,\tau)}{\partial \tau}=\frac{\partial^2 \theta(\eta,\tau)}{\partial \eta^2}-R'\xi(\eta,\tau),
  \label{eq:nond_oxy_final}
\end{equation}
where $R'=\frac{R_0'}{2}(\frac{\theta(\eta,\tau)}{\theta(\eta,\tau)+K_{\mathrm{d1}}'}+\frac{\theta(\eta,\tau)}{\theta(\eta,\tau)+K_{\mathrm{d2}}'})$ with $R_0'=\frac{B_{\mathrm{tot}}h^2 R_0}{c_0D_{\mathrm{oxy}}}$,
$K_{\mathrm{d1}}'=K_{\mathrm{d1}}/c_0$, and $K_{\mathrm{d2}}'=K_{\mathrm{d2}}/c_0$.

Next, to nondimensionalize the equation for bacterial distribution,
we first transform the advection velocities as 
\[
    u_{\mathrm{hydro}}(y,t)=-k(\frac{1}{y^2}-\frac{1}{(h-y)^2})=-\frac{k}{h^2}(\frac{1}{\eta^2}-\frac{1}{(1-\eta)^2})
\] and 
\[
    u_{\mathrm{taxis}}(y,t)=\chi \frac{\partial c(y,t)/\partial y}{(c(y,t)+K_{\mathrm{d1}})(c(y,t)+K_{\mathrm{d2}})}
    =\frac{\chi}{hc_0}\frac{\partial\theta(\eta,\tau)/\partial\eta}{(\theta(\eta,\tau)+K_{\mathrm{d1}}')(\theta(\eta,\tau)+K_{\mathrm{d2}}')}.
\]
Then, Eq. 1 becomes
\begin{equation}
 \frac{\partial \xi(\eta,\tau)}{\partial \tau}=\frac{\partial}{\partial \eta}[d\frac{\partial \xi(\eta,\tau)}{\partial \eta}
  +(k'(\frac{1}{\eta^2}-\frac{1}{(1-\eta)^2})
  -\chi'\frac{\partial\theta(\eta,\tau)/\partial\eta}{(\theta(\eta,\tau)+K_{\mathrm{d1}}')(\theta(\eta,\tau)+K_{\mathrm{d2}}')}
  )\xi(\eta,\tau)] = \frac{\partial \Xi}{\partial \eta},
  \label{eq:nond_bac_final}
\end{equation}
where $k'=k/D_{\mathrm{oxy}}h$ and $\chi'=\chi/D_{\mathrm{oxy}}c_0$ represent the nondimensionalized hydrodynamic strength and aerotactic sensitivity, respectively, with $d=D_{\mathrm{bac}}/D_{\mathrm{oxy}}$.

The boundary conditions for Eq.~\ref{eq:nond_oxy_final} include 
$\left.
\frac{\partial \theta(\eta,\tau)}{\partial \eta}
\right|_{\eta=0}=0$ and $\theta(\eta=1,\tau)=c_h/c_0=c_h'$, in which $c_h'$ is determined by balancing the nondimensionalized total respiration rate and oxygen flux, as explained in the main text.

For Eq.~\ref{eq:nond_bac_final}, we adopt $\left.\Xi\right|_{\eta=0}=0$ and apply the normalization condition $\int_0^1 \xi(\eta)\,\mathrm{d}\eta=1$.

\section{Discussion on model calculations in extended parameter space}

\begin{figure*}[h]
\centering
\includegraphics[width=0.5\textwidth]{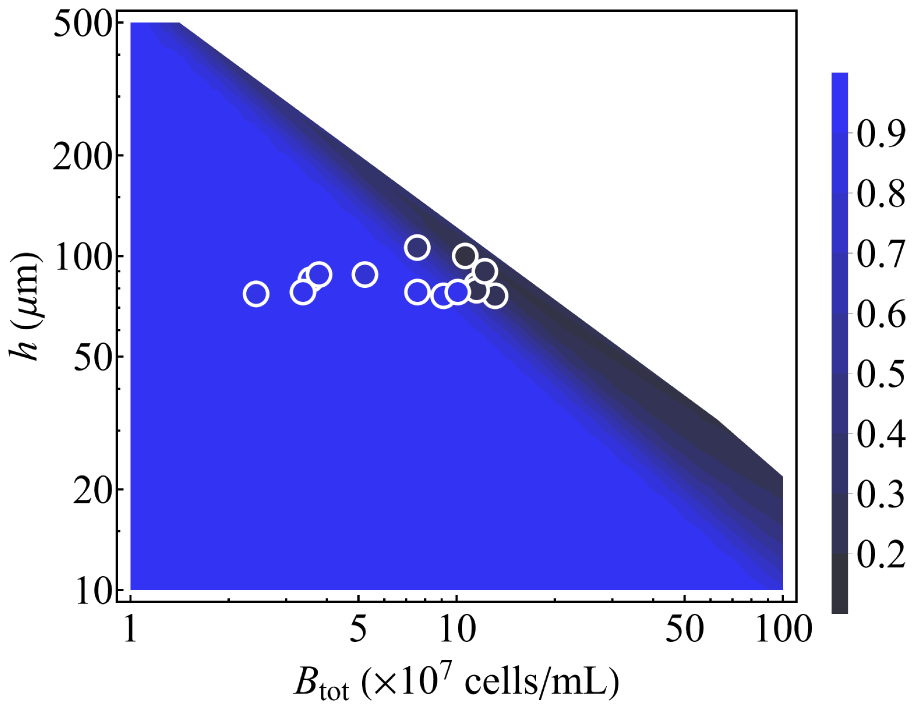}
\caption{
\label{fig:extended model}
The state diagram of bacterial distribution symmetry over a broader range of \(B_{\mathrm{tot}}\) and $h$.
All other details, including the color scale and representation of experimental data, follow the same convention as in the main-text state diagram.
}
\end{figure*}

Fig.~\ref{fig:extended model} shows our model calculations for extended ranges of total bacterial densities \(B_{\mathrm{tot}}\) and film heights $h$. We discuss two extreme regimes in the state diagram where additional physical effects become relevant.

For large $h$ and $B_{\mathrm{tot}}$, one should consider bioconvection: a phenomenon arising from the competition between aerotactic migration and gravitational instability~\cite{Yanaoka2022Bioconvection3D, GallardoNavarro2025Segregation, Shoup2023Bioconvection}.
Specifically, the highly asymmetric bacterial distribution creates a density inversion, i.e., intense bacterial accumulation at the oxygen-supplying top, possibly triggering Rayleigh-Taylor instability.
This causes bacterial plumes to descend toward the bottom, while the vertical oxygen gradient drives them to migrate upward again. 
This interplay generates macroscopic convection rolls, manifesting as various bioconvective patterns depending on film height and bacterial species~\cite{GallardoNavarro2025Segregation,Shoup2023Bioconvection}.
However, in our experiments, we do not observe such bioconvective patterns, even under high $B_{\mathrm{tot}}$ conditions yielding strongly asymmetric density profiles. 
To assess the stability of our system, we calculate the Rayleigh number ($Ra = \Delta\rho g h^3 / D_{\mathrm{bac}}\mu$), a dimensionless number characterizing the onset of bioconvection.
$Ra$ in previous experimental studies reporting robust bioconvection is typically of the order of thousands, while numerical simulations suggest a critical value of $Ra_{\mathrm{crit}} \approx 500$~\cite{Yanaoka2022Bioconvection3D}.
The maximum estimated Rayleigh number in our experimental setup is $Ra \approx 5$, which is well below $Ra_{\mathrm{crit}}$, supporting our observation that the system remains stable with no bioconvective instabilities.

As $h$ decreases, wall accumulation may weaken, and bulk concentration increases~\cite{Wei2025Confinement}.
This phenomenon is attributed to the hydrodynamic quadrupole flow, which exerts a torque that reorients bacteria toward the center of the film, thereby reducing accumulation at the boundaries.
This bulk-concentrating effect becomes notable for heights $h \le 30$ \si{\um}, a range that falls outside our experimental conditions.
Note, however, that our model calculations in Fig.~\ref{fig:extended model} are based on dipoles only and predict persistent wall accumulation even in this regime.
Further investigation into quadrupole corrections and experimental exploration of extreme confinement is warranted.

\FloatBarrier

%
